\documentclass[aps,pra,reprint,groupedaddress,showkeys]{revtex4-1}
\usepackage{amsmath,graphicx,hyperref,verbatimbox}
\usepackage[FIGTOPCAP,raggedright,nooneline,bf,footnotesize]{subfigure}
\usepackage{indentfirst}
\usepackage{braket}
\usepackage{diagbox}
\usepackage{amssymb}
\usepackage{bbold}
\usepackage{colortbl}
\usepackage{multirow}

\newcommand{\figref}[1]{Fig.~\ref{#1}}

\renewcommand{\eqref}[1]{Eq.~\ref{#1}}
\newcommand{\tr}{{\mathrm{Tr}}}

\begin{document}

\title{Quantum State Tomography of Four-Level Systems with Noisy Measurements}
\author{Artur Czerwinski \hspace{0.001cm}}
\email{aczerwin@umk.pl}
\affiliation{Institute of Physics, Faculty of Physics, Astronomy and Informatics \\ Nicolaus Copernicus University,
Grudziadzka 5, 87--100 Torun, Poland}

\begin{abstract}
In this article, we investigate the problem of state reconstruction of four-level quantum systems. A realistic scenario is considered with measurement results distorted by random unitary operators. Two frames which define injective measurements are applied and compared. By introducing arbitrary rotations, we can test the performance of the framework versus the amount of experimental noise. The results of numerical simulations are depicted on graphs and discussed. In particular, a class of entangled states is reconstructed. The concurrence is used as a figure of merit in order to quantify how well entanglement is preserved through noisy measurements.
\end{abstract}
\maketitle

\section{Introduction}

The problem of complex vector reconstruction (phase retrieval) appears in many areas of research \cite{Cahill2015,Jaganathan2016}. The goal is to uniquely determine an unknown vector $\ket{x} \in \mathbb{C}^d$ based on modulus of inner product of the vector in question with elements of a frame $\Xi =\{\ket{\xi_1}, \dots, \ket{\xi_n} \}$. Thus, the accessible data, referred to as intensity measurements, can be written in the form:
\begin{equation}\label{i1}
\left|\braket{\xi_i | x}\right|^2 \hspace{0.5cm}\text{for}\hspace{0.5cm} i=1,\dots, n.
\end{equation}
If phase retrieval is possible, we say that the frame $\Xi$ defines injective measurements. So far sufficient conditions have been formulated for the injectivity of measurements. For example, it has been proved that a generic frame with the number of elements satisfying $n \geq 4 (d-1)$ is sufficient to reconstruct an unknown complex vector $\ket{x}\in\mathbb{C}^d$ \cite{bandeiraa14,conca15}. However, necessary conditions for phase retrieval still remain obscure and the figure $4 (d-1)$ cannot be considered a threshold expressing the minimal number of measurements. In particular, for $\ket{x}\in\mathbb{C}^4$ it has been demonstrated that a frame consisting of $11$ elements defines injective measurements \cite{vinzant15}. In the present article, this specific frame is applied and tested in an imperfect measurements scenario.

In quantum physics, the problem of phase retrieval appears if we want to reconstruct a complex vector $\ket{\psi} \in \mathcal{H}$ which represents a pure state of a physical system. This question belongs to a subfield called \textit{quantum state tomography} (QST), which aims at recovering accurate mathematical representations of quantum states from measurements \cite{dariano03,paris04}. Some proposals, both theoretical and experimental, focus on performing QST with the minimal number of measurements \cite{Oren2017,Martinez2019}. Usually, the post-measurement state of the system is of little interest since the probabilities of the respective measurement outcomes are in the center of attention. In such cases, positive operator-valued measures (POVMs) can be applied to study the statistics of measurements \cite{Nielsen2000}. In particular, symmetric informationally complete POVMs (SIC-POVMs) can be considered optimal as far as the number of measurements is concerned \cite{Rehacek2004,Renes2004,Fuchs2017}. Special attention should be paid to the methods which utilize dynamical maps in order to decrease the number of necessary measurement operators \cite{Czerwinski2016a,Czerwinski2020a}. On the other hand, in practical realizations of QST protocols, there is a tendency to apply overcomplete sets of measurements in order to reduce the detrimental impact of experimental noise \cite{Horn2013,Zhu2014}. Particularly, mutually unbiased bases (MUBs) can be employed as an overcomplete measurement scheme \cite{Wootters1989,Durt2010}.

In this article, we consider only a finite-dimensional Hilbert space $\mathcal{H} \cong \mathbb{C}^d$, i.e. $\dim \mathcal{H} = d < \infty$. More specifically, we investigate the problem of QST of four-level systems described by pure states. Reconstruction of pure states (or almost pure) usually involves separate tomographic techniques \cite{Gross2010,Bantysh2020,Zambrano2020}. In our scheme, two generic frames are applied in order to reconstruct a sample of $4-$dimensional complex vectors. One frame comprises $20$ vectors which belong to the MUBs for the Hilbert space $\mathcal{H}$ such that $\dim \mathcal{H} = 4$. The other one, introduced by C.~Vinzant \cite{vinzant15}, consists of $11$ elements. The connection between QST and the theory of frames has already been studied in the context of the stroboscopic approach to quantum state identification \cite{Jamiolkowski2010,Jamiolkowski2012}.

From a physical point of view, intensity measurements of the form \eqref{i1} correspond to an unnormalized POVM since:
\begin{equation}\label{i2}
\left|\braket{\xi_i | \psi }\right|^2 = \tr \left( \ket{\psi} \! \bra{\psi} M_i \right),
\end{equation}
where the measurement operator $M_i$ is defined as a rank-one projector, i.e. $M_i := \ket{\xi_i} \! \bra{\xi_i}$. Assuming that we can normalize the frame vectors, we obtain measurement results which are equivalent to probabilities given by the Born's rule. Therefore, the kind of measurement analyzed in the paper is in line with the general description of quantum measurement.

This work is a follow-up of the article \cite{Czerwinski2020b}, where qubits were reconstructed with two frames comprising distinct numbers of elements. However, in the present article we study a different source of experimental noise. We impose random rotations on the measurement operators and investigate the efficiency of the frames in quantum state reconstruction for various degrees of experimental noise. Moreover, since four-level systems are considered, we can analyze QST of entangled states, as a specific example. The results indicate how well entanglement can be retrieved from imperfect measurements.

In Sec.~\ref{method}, we introduce the framework of QST of four-level systems along with the figures of merit which are used to quantify the efficiency of our tomographic scheme. Then, in Sec.~\ref{results}, we present and discuss the results of numerical simulations. The figures of merit are depicted on graphs, which allows one to observe how the efficiency of the QST framework depends on the amount of experimental noise. The framework can be successfully applied to study QST of pure states with different frames.

\section{State reconstruction with noisy measurements}\label{method}

In this work, we assume that the initial state of a four-level system can be presented as a complex vector:
\begin{equation}\label{m1}
\ket{\psi_{in}} =\begin{pmatrix} \cos \frac{\theta}{2} \sin \frac{\beta}{2}  \\ 
\\
 \sin \frac{\theta}{2} \sin \frac{\beta}{2} \: e^{i \phi_{12}}\\
 \\
 \sin \frac{\delta}{2} \cos \frac{\beta}{2}  \: e^{i \phi_{13}}\\
 \\
  \cos \frac{\delta}{2}  \cos \frac{\beta}{2} \: e^{i \phi_{14}} \end{pmatrix},
\end{equation}
where $ 0\leq \phi_{12}, \phi_{13}, \phi_{14} < 2 \pi$ and $0\leq \theta, \beta, \delta \leq \pi$. The parametrization \eqref{m1} represents a general $4-$level pure state, where $\phi_{12}, \phi_{13}, \phi_{14}$ denote relative phases between the respective basis states. An unknown quantum state of the form \eqref{m1} can be reconstructed from injective measurements generated by a generic frame: $\Xi = \{ \ket{\xi_1}, \dots\}$, where $\ket{\xi_i} \in \mathbb{C}^4$. In the context of physical applications, we assume that the frame vectors are normalized, which implies that the intensity measurements $\left|\braket{\xi_i | \psi_{in}}\right|^2$ are equivalent to probabilities in the projective measurement.

To make the framework realistic, we assume that the measurements are subject to experimental noise, which can be mathematically modeled by random unitary transformations that distort the original frame vectors, see \cite{Lohani2020}. The general form a $2 \times 2$ unitary rotational operator is given by:
\begin{equation}\label{m2}
U (\omega_1, \omega_2, \omega_3) = \begin{pmatrix} e ^{i \omega_1/ 2 } \cos \omega_3 & & -i e^{i \omega_2} \sin \omega_3 \\ & & \\ - i e^{-i \omega_2} \sin \omega_3 & & e^{- i \omega_1/2} \cos \omega_3  \end{pmatrix},
\end{equation}
where the parameters: $\omega_1, \omega_2, \omega_3$, in our application, are selected randomly from a normal distribution characterized by the mean value equal $0$ and a non-zero standard deviation denoted by $\sigma$, i.e. $\omega_1, \omega_2, \omega_3 \in \mathcal{N}(0,\sigma)$. This allows us to construct a $4\times4$ perturbation matrix $\mathcal{P} (\sigma)$ as:
\begin{equation}\label{m3}
\mathcal{P} (\sigma) := U (\omega_1, \omega_2, \omega_3) \otimes U (\omega'_1, \omega'_2, \omega'_3).
\end{equation}
Equipped with the definition \eqref{m3}, we can introduce a simulated result of the $k-$th intensity measurement burdened with experimental noise:
\begin{equation}\label{m4}
p^{M}_k = \left|\braket{ \xi_k (\sigma) | \psi_{in}}\right|^2,
\end{equation}
where $\ket{\xi_k (\sigma)} = \mathcal{P}(\sigma) \ket{\xi_k}$. For each single measurement, a different perturbation matrix $\mathcal{P} (\sigma)$ is generated with random parameters according to \eqref{m3} and \eqref{m2}, which allows us to obtain noisy measurement results with a given parameter $\sigma$. Thanks to this approach, each act of observation is burdened with random uncertainty and $\sigma$ is used to quantify the amount of experimental noise.

For a specific frame $\Xi$, we are able to numerically generate experimental data corresponding to any input state of the form \eqref{m1}. However, when we reconstruct an unknown state of a quantum system, we assume that the experimenter does not possess any a priori knowledge about the state in question. Thus, we utilize the Cholesky factorization, cf. Ref.~\cite{James2001,Altepeter2005,SedziakKacprowicz2020}, which gives a general representation of $4\times4$ density matrix:
\begin{equation}\label{m5}
\rho_{out} (t_1, \dots, t_{16}) = \frac{T^{\dagger} T}{\tr\: \left(T^{\dagger} T\right)},
\end{equation}
where:
\begin{equation}\label{m6}
T=\begin{pmatrix} t_1 & 0 & 0 &0 \\ t_5 + i\, t_6 & t_2 & 0 &0 \\  t_{11} + i \,t_{12} & t_7 + i\, t_8 & t_3 &0 \\ t_{15} + i\, t_{16} & t_{13} + i\, t_{14} & t_9 + i\, t_{10} & t_4 \end{pmatrix},
\end{equation}
which means that we need to estimate the values of $16$ real parameters: $t_1, t_2, \dots, t_{16}$ in order to obtain the complete knowledge about an unknown state. Thanks to the Cholesky decomposition, any density matrix resulting from the framework is physical, i.e. it is Hermitian, positive semi-definite, of trace one.

With $\rho_{out}$ standing for the output density matrix, we can write a formula, according to the Born's rule, for the expected result of $k-$th measurements:
\begin{equation}\label{m7}
p^{E}_k = \tr \left( \ket{\xi_k}\! \bra{\xi_k} \, \rho_{out} (t_1, \dots, t_{16}) \right).
\end{equation}
In order to determine the values of the parameters $t_1, \dots, t_{16}$ that fit optimally to the noisy measurements, we shall apply the method of least squares (LS) \cite{Opatrny1997}. This techniques, together with the maximum likelihood estimation (MLE), is often implemented in different tomographic frameworks, see e.g. Ref.~\cite{Acharya2019,SedziakKacprowicz2020}. The LS method requires to search for the minimum value of the following function:
\begin{equation}\label{m8}
\begin{aligned}
{}&f^{LS}_{\sigma} (t_1, t_2, \dots, t_{16}) = \sum_k \left( p^{E}_k  - p^{M}_k  \right)^2 = \\& =\sum_k \left(  \tr \left( \ket{\xi_k}\! \bra{\xi_k}  \, \rho_{out}(t_1, \dots, t_{16})  \right) - \left|\braket{ \xi_k (\sigma) | \psi_{in}}\right|^2 \right)^2,
\end{aligned}
\end{equation}
which can be done numerically for any initial state $\ket{\psi_{in}}$ and a specific frame $\Xi$.

In the present work, we compare the efficiency of two frames in state tomography of four-level systems. The quality of state reconstruction is quantified by two figures of merit: quantum fidelity, $\mathcal{F} (\sigma)$, given by \cite{Nielsen2000}:
\begin{equation}\label{m9}
\mathcal{F} (\sigma) := \left(\tr \sqrt{\sqrt{\rho_{out}}\ket{\psi_{in}} \!\bra{\psi_{in}} \sqrt{\rho_{out}}} \right)^2
\end{equation}
and purity, $\gamma (\sigma)$, defined as:
\begin{equation}
\gamma (\sigma) :=  \tr \left( \rho_{out}^2 \right).
\end{equation}
In our framework, both figures depend on the amount of noise introduced into the measurements. For two quantum states, the fidelity measures their closeness \cite{Uhlmann1986,Jozsa1994}, whereas the purity is commonly used to evaluate how far the state has drifted from the pure. Both quantities can be used to compare the result of a QST framework with the original. In addition, they can be applied to track changes that occur to quantum state in time \cite{Czerwinski2020d}.

In our model, we perform QST with each frame for a sample of $12\,096$ input states defined as \eqref{m1}, with the parameters $ \phi_{12}, \phi_{13}, \phi_{14}, \theta,\beta, \delta$ covering the full range (discretely selected with a proper step). Then, the performance of each frame can be expressed by the average fidelity $\mathcal{F}_{av} (\sigma)$ and purity $\gamma_{av} (\sigma)$ computed over the sample. A similar approach to evaluate the performance of quantum state estimation was utilized in \cite{SedziakKacprowicz2020,Czerwinski2020b}.

Apart from a general sample of $12\,096$ four-level input states, we shall consider a special subset of the states given by:
\begin{equation}\label{m10}
\ket{\Phi_{in}} = \frac{1}{\sqrt{2}} \left(\ket{00} + e^{i \phi} \ket{11} \right),
\end{equation}
where $\ket{00}$ and $\ket{11}$ denote two vectors from the standard basis in $\mathbb{C}^4$, whereas $\phi$ represents the relative phase between the states ($0\leq \phi < 2 \pi)$. For two specific values of the relative phase, i.e. $\phi = 0$ or $\phi = \pi$, one gets the elements of the Bell basis, denoted by $\Phi^+$ and $\Phi^-$, respectively.

This particular state vector \eqref{m10} describes one type of two-qubit entanglement, which can be realized on photons by exploiting different degrees of freedom, especially: polarization, spectral, spatial and temporal mode. Such states are commonly considered in quantum communication protocols \cite{Wang2009,Herbauts2013}, because one can generate this kind of entanglement by a variety of experimental techniques, for example by spontaneous four-wave mixing (SFWM) in a dispersion shifted fiber \cite{Takesue2005,Takesue2009}, or by spontaneous parametric down conversion (SPDC) \cite{Marcikic2004,Shimizu2009}, and by a source which utilize quantum dots \cite{Jayakumar2014,Versteegh2015}.

The problem of relative phase estimation for quantum states of the form \eqref{m10} was first solved for polarization entangled photons \cite{White1999}. Recently, such states have been considered in a QST framework devoted to time-bin qudits \cite{SedziakKacprowicz2020}.

In our framework, we analyze, as a separate case, the performance of the frames in QST of states given by \eqref{m10} by computing the average fidelity for a sample of $200$ entangled states (selected for the whole spectrum of $\phi$). Additionally, in order to measure how much entanglement is retrieved from the measurements, we compute the concurrence for each density matrix $\rho_{out}$ obtained from the QST scheme \cite{Hill1997,Wootters1998}. First, we obtain the spin-flipped state $\tilde{\rho}_{out}$ which is defined as:
\begin{equation}\label{m11}
\tilde{\rho}_{out} = \left(\sigma_y \otimes \sigma_y \right) \rho^*_{out} \left(\sigma_y \otimes \sigma_y \right),
\end{equation}
where $\rho^*_{out}$ stands for the complex conjugate (provided we operate in the standard basis) and $\sigma_y$ denotes one of the Pauli spin matrices, i.e. $\sigma_y = \begin{pmatrix} 0 & -i \\ i & 0 \end{pmatrix}$. Then, the $R-$matrix is built by the formula:
\begin{equation}\label{m12}
R := \sqrt{\sqrt{\rho_{out}} \,\tilde{\rho}_{out}\, \sqrt{\rho_{out}}},
\end{equation}
which allows us to define the concurrence, $C(\rho_{out})$, by means of the eigenvalues of the $R-$matrix:
\begin{equation}\label{m13}
C(\rho_{out}) := \max \left\{0, \alpha_1 - \alpha_2 - \alpha_3 -\alpha_4   \right\},
\end{equation}
where $\alpha_1, \alpha_2, \alpha_3, \alpha_4$ are the eigenvalues of the $R-$matrix in the decreasing order.

For any density matrix $\rho$, the concurrence satisfies: $0\leq C(\rho) \leq 1$. We have $C(\rho) = 1$ for maximally entangled states and $C(\rho) = 0$ for separate states. Thus, the concurrence can be considered an entanglement monotone, which means it can be applied to quantify quantum entanglement, see \cite{Buchleitner2007}.

Finally, as another figure of merit, we use the average concurrence computed over the sample, $C_{av} (\sigma)$, which is presented on graphs as a function of the amount of noise. 

\section{Results and discussion}\label{results}

In the main part of the article, we compare the efficiency of two frames in QST of pure states which describe four-level quantum systems. The frames differ in the number of elements and for this reason were selected as the case study.

The first frame, denoted by $Z^{MUB}$, consists of $20$ vectors which correspond to the elements of the MUBs for $\dim \mathcal{H} =4$ \cite{Klappenecker2004}, i.e.:
\begin{equation}\label{r1}
\begin{split}
&\ket{\zeta_1} = \begin{pmatrix} 1 \\ 0 \\ 0 \\ 0 \end{pmatrix}, \hspace{0.15cm} \ket{\zeta_2} = \begin{pmatrix} 0  \\ 1 \\ 0 \\0 \end{pmatrix}, \hspace{0.15cm} \ket{\zeta_3} = \begin{pmatrix} 0  \\ 0 \\ 1 \\0 \end{pmatrix}, \hspace{0.15cm} \ket{\zeta_4} = \begin{pmatrix} 0  \\ 0 \\ 0 \\1 \end{pmatrix},\\
&\ket{\zeta_5} = \frac{1}{2}\begin{pmatrix} 1 \\ 1 \\ 1 \\ 1 \end{pmatrix}, \hspace{0.5cm} \ket{\zeta_6} = \frac{1}{2}\begin{pmatrix} 1  \\ 1 \\ -1 \\-1 \end{pmatrix}, \hspace{0.15cm} \ket{\zeta_7} = \frac{1}{2}\begin{pmatrix} 1  \\-1 \\ -1 \\1 \end{pmatrix},\\
& \ket{\zeta_8} = \frac{1}{2}\begin{pmatrix} 1  \\ -1 \\ 1 \\-1 \end{pmatrix},\hspace{0.15cm} \ket{\zeta_9} = \frac{1}{2}\begin{pmatrix} 1 \\ -1 \\ -i \\ -i \end{pmatrix}, \hspace{0.15cm} \ket{\zeta_{10}} = \frac{1}{2}\begin{pmatrix} 1  \\ -1 \\ i \\i \end{pmatrix},\\
&\ket{\zeta_{11}} = \frac{1}{2}\begin{pmatrix} 1  \\1 \\ i \\ -i \end{pmatrix},\hspace{0.15cm} \ket{\zeta_{12}} = \frac{1}{2}\begin{pmatrix} 1  \\ 1 \\ -i \\ i \end{pmatrix}, \hspace{0.15cm} \ket{\zeta_{13}} = \frac{1}{2}\begin{pmatrix} 1 \\ -i \\ -i \\ -1 \end{pmatrix},\\
&\ket{\zeta_{14}} = \frac{1}{2}\begin{pmatrix} 1  \\ -i \\ i \\1 \end{pmatrix}, \hspace{0.15cm} \ket{\zeta_{15}} = \frac{1}{2}\begin{pmatrix} 1  \\i \\ i \\-1 \end{pmatrix},\hspace{0.15cm} \ket{\zeta_{16}} = \frac{1}{2}\begin{pmatrix} 1  \\ i \\ -i \\1 \end{pmatrix},\\
&\ket{\zeta_{17}} = \frac{1}{2}\begin{pmatrix} 1 \\ -i \\ -1 \\ -i \end{pmatrix}, \hspace{0.15cm} \ket{\zeta_{18}} = \frac{1}{2}\begin{pmatrix} 1  \\ -i \\ 1 \\ i \end{pmatrix}, \hspace{0.15cm} \ket{\zeta_{19}} = \frac{1}{2}\begin{pmatrix} 1  \\i \\ -1 \\i \end{pmatrix},\\
& \ket{\zeta_{20}} = \frac{1}{2}\begin{pmatrix} 1  \\ i \\ 1 \\-i \end{pmatrix}.\\
\end{split}
\end{equation}
The frame $Z^{MUB} = \{ \zeta_1, \zeta_2, \dots, \zeta_{20}\}$ defines injective measurements which can be considered overcomplete. We confront this frame with a minimal set of intensity measurements introduced by C.~Vinzant \cite{vinzant15}. The minimal frame, which shall be denoted by $\Lambda^{MIN}$, comprises $11$ vectors:
\begin{equation}\label{r2}
\begin{split}
&\ket{\lambda_1} = \begin{pmatrix} 1 \\ 0 \\ 0 \\ 0 \end{pmatrix}, \hspace{0.15cm} \ket{\lambda_2} = \begin{pmatrix} 0  \\ 1 \\ 0 \\0 \end{pmatrix}, \hspace{0.15cm} \ket{\lambda_3} = \begin{pmatrix} 0  \\ 0 \\ 1 \\0 \end{pmatrix}, \hspace{0.15cm} \ket{\lambda_4} = \begin{pmatrix} 0  \\ 0 \\ 0 \\1 \end{pmatrix},\\
&\ket{\lambda_5} = \begin{pmatrix} 1 \\ 9 i \\ -5 - 7 i \\ -6- 7i \end{pmatrix}, \hspace{0.5cm} \ket{\lambda_6} = \begin{pmatrix} 1  \\ 1-i \\ -5-2 i \\-1-8i \end{pmatrix},\\
& \ket{\lambda_7} = \begin{pmatrix} 1  \\-2+ 4 i \\ -4 - 2 i \\ 3 + 8 i \end{pmatrix},\hspace{0.5cm} \ket{\lambda_8} =\begin{pmatrix} 1  \\ -3+ i \\ 1-8 i \\7-6 i \end{pmatrix},\\
&\ket{\lambda_9} = \begin{pmatrix} 1 \\ 3-3i \\ -8+7i \\ -6-2 i \end{pmatrix}, \hspace{0.5cm} \ket{\lambda_{10}} =\begin{pmatrix} 1 \\ -3 + 5 i \\ 5+ 6 i \\ 2 i \end{pmatrix},\\
&\ket{\lambda_{11}} =\begin{pmatrix} 1  \\ -3 + 8 i \\ 5 - 5 i  \\ -6 - 4 i \end{pmatrix},
\end{split}
\end{equation}
which are also sufficient to recover any unknown vector $\ket{x} \in \mathbb{C}^4$. For the sake of physical rigor, before applying the method of least squares, we normalize the frame vectors, i.e. $\ket{\tilde{\lambda}_k} = \frac{\ket{\lambda_k}}{\sqrt{\braket{\lambda_k|\lambda_k}}}$, but this operation does not change the algebraic properties of the frame.

In order to investigate the efficiency of each frame in four-level state reconstruction, numerical simulations were conducted, assuming different values of the standard deviation $\sigma$, which governs the experimental noise according to \eqref{m4}. A sample of $12\,096$ input states of the form \eqref{m1} was considered and each state was reconstructed with measurements distorted by the random unitary rotation operators.

\begin{figure}[h]
	\centering
   \begin{subfigure}
         \centering
         \includegraphics[width=0.95\columnwidth]{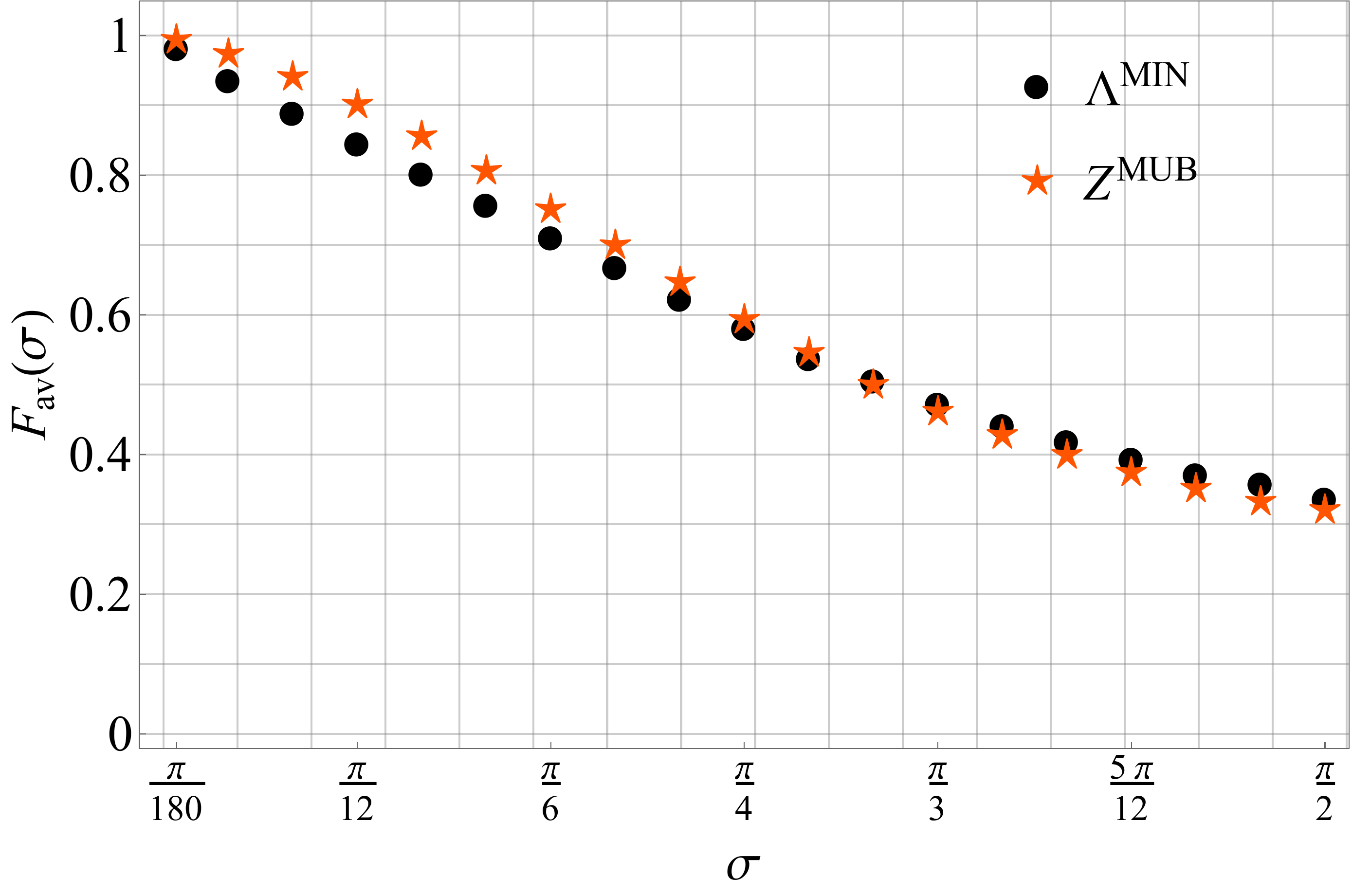}
     \end{subfigure}
     \hfill
     \begin{subfigure}
         \centering
         \includegraphics[width=0.95\columnwidth]{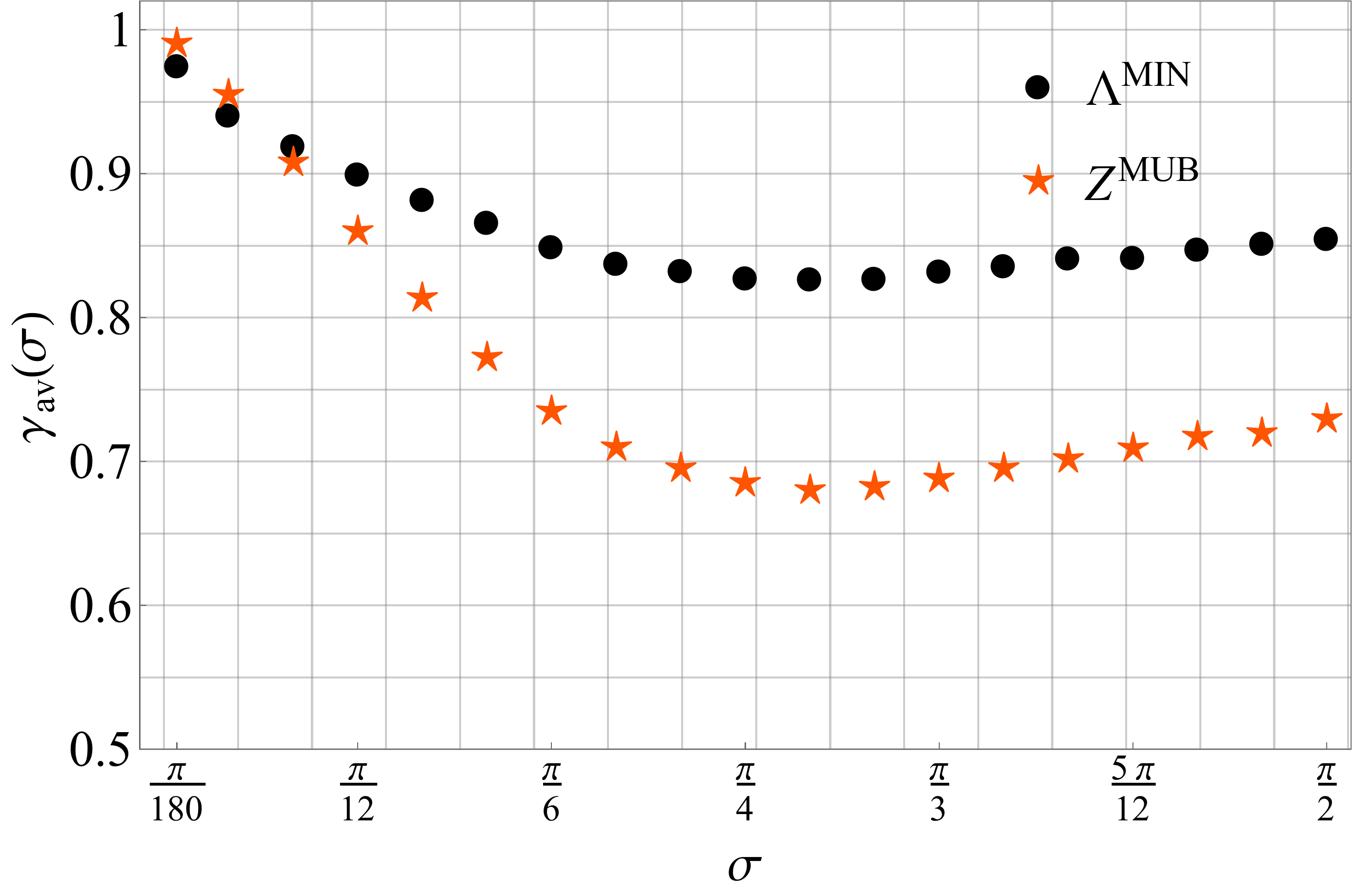}
     \end{subfigure}
	\caption{Plots present the average fidelity $\mathcal{F}_{av} (\sigma)$ (the upper graph) and the purity $\gamma_{av} (\sigma)$ (the lower graph) in QST of four-level systems with the frames $Z^{MUB}$ and $\Lambda^{MIN}$. Each point was obtained by the method of least squares for a sample of $12\,096$ input states of the form \eqref{m1}. The formula \eqref{m4} was applied for the measurement results with the experimental noise governed by the standard deviation $\sigma$.}
	\label{plots1}
\end{figure}

In \figref{plots1}, one can observe the plots of the average fidelity $\mathcal{F}_{av} (\sigma)$ and the purity $\gamma_{av} (\sigma)$ for a sample of four-level states. The average fidelity, which gives the overlap between the actual state and the result of QST, is a crucial indicator of the accuracy of state reconstruction. One can notice that the overcomple frame, $Z^{MUB}$, has a modest advantage over the minimal frame, $\Lambda^{MIN}$. This result appears rather surprising since the frames differ by $9$ elements and one would expect more significant discrepancies in the performance of the frames.

Only for moderate degrees of noise, i.e. as long as $\sigma < 0.9$, the average fidelity corresponding to the frame $Z^{MUB}$ is greater than the one resulting from applying $\Lambda^{MIN}$. Interestingly, for highest degrees of noise (i.e. $\sigma > \pi/3$), the performance of $\Lambda^{MIN}$ in the QST framework is slightly better than $Z^{MUB}$.

The results seem even more intriguing if we investigate the average purity, $\gamma_{av} (\sigma)$, presented in \figref{plots1} (the lower graph). It turns out that for $\sigma \geq \pi /18$ the average purity of the states $\rho_{out}$ obtained from the frame $\Lambda^{MIN}$ is greater than $Z^{MUB}$. Both plots included in \figref{plots1} demonstrate that the overcomplete frame $Z^{MUB}$ is advantageous only for a little amount of noise.

\begin{figure}[h]
	\centering
   \begin{subfigure}
         \centering
         \includegraphics[width=0.95\columnwidth]{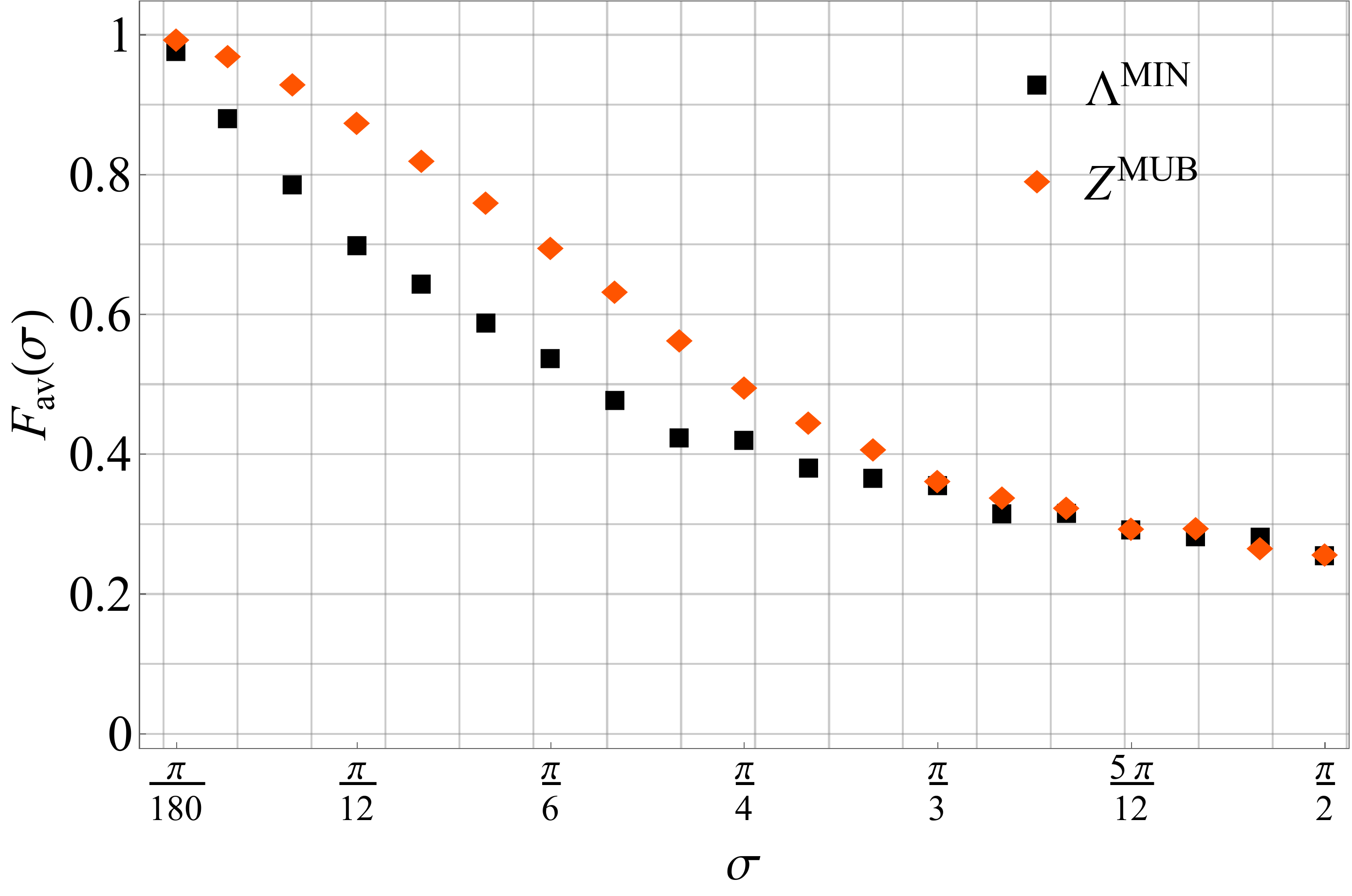}
     \end{subfigure}
     \hfill
     \begin{subfigure}
         \centering
         \includegraphics[width=0.95\columnwidth]{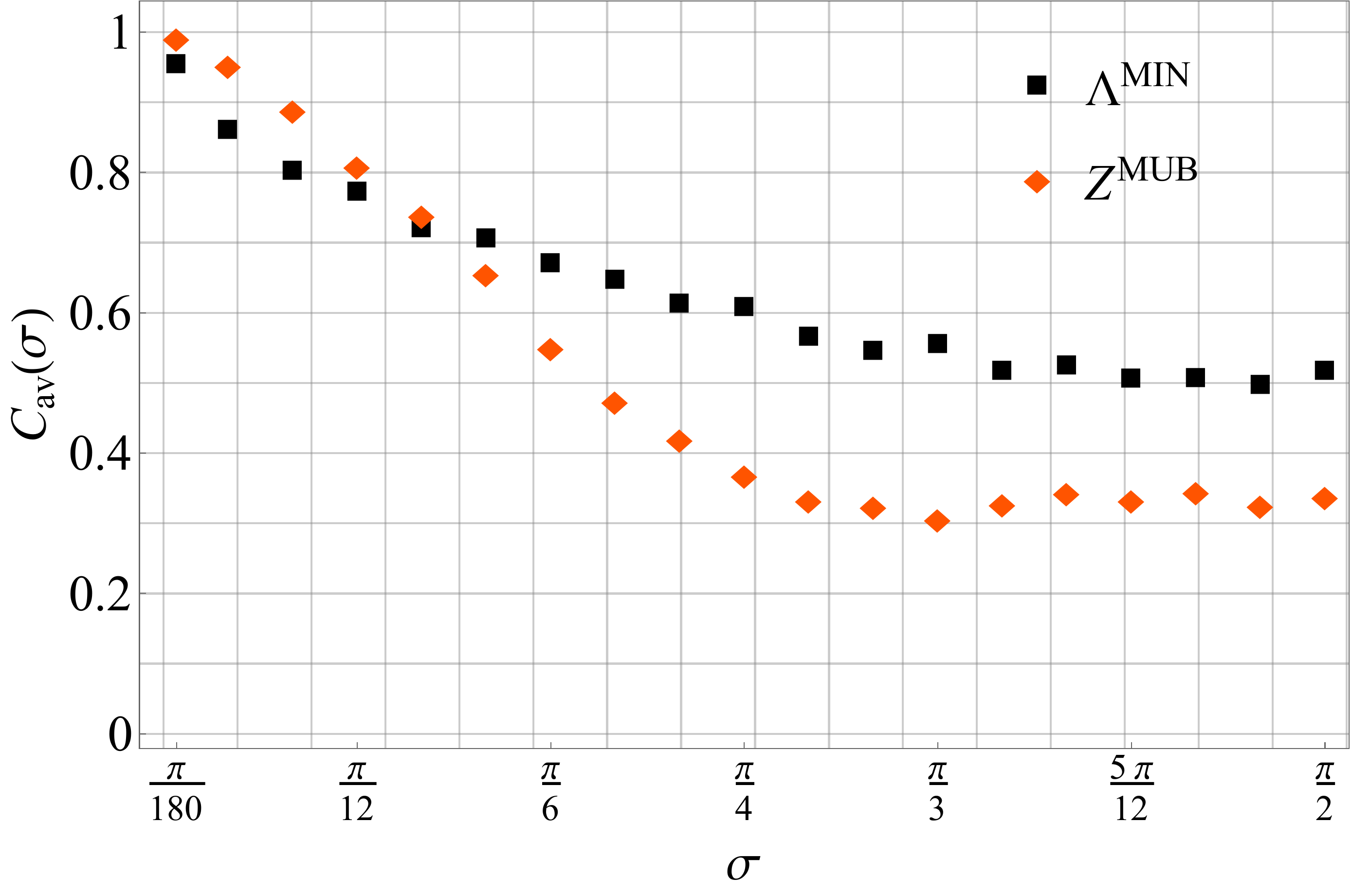}
     \end{subfigure}
	\caption{Plots present the average fidelity $\mathcal{F}_{av} (\sigma)$ (the upper graph) and the concurrence $C_{av} (\sigma)$ (the lower graph) in QST of entangled qubits with the frames $Z^{MUB}$ and $\Lambda^{MIN}$. Each point was obtained by the method of least squares for a sample of $200$ input states of the form \eqref{m10}. The formula \eqref{m4} was applied for the measurement results with the experimental noise governed by the standard deviation $\sigma$.}
	\label{plots2}
\end{figure}

Next, we consider QST of input states in the form \eqref{m10}. Although it is a class of entangled states which differ only in the relative phase $\phi$, it is worth stressing that we still follow the general formula for an unknown density matrix \eqref{m5}. This means that we select a sample of $200$ states and in each case we estimate the values of $16$ parameters which completely characterize four-level state. In \figref{plots2}, one can observe the results of numerical simulations for a wide range of the standard deviation $\sigma$, which describes the degree of experimental noise.

One can notice a substantial advantage of $Z^{MUB}$ over $\Lambda^{MIN}$ as long as the average fidelity is concerned. The frame $Z^{MUB}$ outperforms $\Lambda^{MIN}$ up to $\sigma = \pi/3$, where both plots converge.

The lower graph in \figref{plots2} presents the average concurrence $C_{av} (\sigma)$. Interestingly, if $\sigma > \pi/9$, the minimal frame $\Lambda^{MIN}$ leads to quantum states which feature a greater amount of entanglement than the results stemming from $Z^{MUB}$. However, in most applications, we are interested in detecting entangled states sufficient to announce the violation of the Bell inequality \cite{Zukowski2002}. This can be guaranteed if the concurrence satisfies $C(\rho) > 1/\sqrt{2}$ \cite{Verstraete2002,Hu2012}. From \figref{plots2}, we can conclude that for $\sigma \leq \pi/9$, with either of the frames, we obtain such quantum states that $C_{av} (\sigma) > 1/\sqrt{2}$. Thus, the interval $0<\sigma \leq \pi/9$ can be considered the allowable noise range which does not disturb the detection of entanglement. Finally, we should stress that for $0<\sigma \leq \pi/9$ the frame $Z^{MUB}$ provides better accuracy in terms of both the average fidelity and concurrence, which proves dominance of the overcomplete frame within the relevant noise interval.

\section{Summary and outlook}

Two generic frames were implemented in QST of four-level pure states. One frame, $Z^{MUB}$, was composed of the elements of the MUBs in the $4-$dimensional Hilbert space, whereas the other, $\Lambda^{MIN}$, comprised $11$ vectors sufficient for phase retrieval. In order to test the performance of the frames, we introduced experimental noise into the measurements by random unitary operators. For a representative sample of four-level states, it was discovered that $Z^{MUB}$ has a moderate advantage over the minimal frame $\Lambda^{MIN}$.

As a special case, we investigated QST of two-qubit entangled states. First, we determined the allowable noise range which guarantees the detection of entanglement. Then, it was demonstrated that within this range $Z^{MUB}$ significantly outperforms $\Lambda^{MIN}$, which proves that overcomplete frames can be more beneficial in QST.

In the future, other classes of entangled states (e.g. entangled qutrits) can be reconstructed with different frames. The framework can also be extended by including additional types of experimental noise. 

\section*{Acknowledgments}

The author acknowledges financial support from the Foundation for Polish Science (FNP) (project First Team co-financed by the European Union under the European Regional Development Fund).

\end{document}